\documentclass[aps,pre,reprint,twocolumn,superscriptaddress,floatfix]{revtex4-1}

\usepackage{graphicx}
\usepackage{amsmath}
\usepackage{amssymb}
\usepackage{epsf}
\usepackage{color}

\begin{document}

\title{Stochastic study of information transmission and population 
stability in a generic bacterial two-component system}

\author{Tarunendu Mapder}
\email{mtarunendu@yahoo.com}
\affiliation{Department of Chemistry, 
Indian Institute of Engineering Science and Technology,
Shibpur, Howrah 711103, India}

\author{Sudip Chattopadhyay}
\email{sudip@chem.iiests.ac.in}
\affiliation{Department of Chemistry, 
Indian Institute of Engineering Science and Technology,
Shibpur, Howrah 711103, India}

\author{Suman K. Banik}
\email{skbanik@jcbose.ac.in}
\affiliation{Department of Chemistry, Bose Institute, 
93/1 A P C Road, Kolkata 700009, India}

\date{\today}

\begin{abstract}
Studies on the role of fluctuations in signal propagation and on gene regulation in monoclonal bacterial population have been extensively pursued based on the machinery of two-component system. The bacterial two-component system shows noise utilisation through its inherent plasticity. The fluctuations propagation takes place using the phosphotransfer module and the feedback mechanism during gene regulation. To delicately observe the noisy kinetics the generic cascade needs stochastic investigation at the mRNA and protein levels. To this end, we propose a theoretical framework to investigate the noisy signal transduction in a generic bifunctional two-component system.
The model shows reliability in information transmission through quantification of several statistical measures. We further extend our analysis to observe the protein distribution in a population of 
cells. Through numerical simulation, we identify the regime of the kinetic parameter set that generates a stability switch in the steady state distribution of proteins.
The results of our theoretical analysis show key features of the network. The noise permeation and information propagation in the autoregulation module is feeble. However, the phosphotransfer module compensates such weakness and plays a significant role in information transmission. The bimodality due to fluctuations pampers the emergence of persistence in an isogenic bacterial pool.
\end{abstract}


\maketitle

\section{Introduction}

To deal with fluctuations in biological systems is highly motivating 
because nature is fundamentally noisy in nature 
\cite{Balazsi2011,Raj2008,Eldar2010}. The interplay 
between deterministic and random processes configures the pool 
of different biological entities along with the adoption of several 
survival strategies. Also, the phenotypic heterogeneity brings 
in uniqueness to an individual in a population \cite{Davidson2008}. 
This happens not only 
for differences at the genetic level but due to the presence of noise 
in an isogenic population. Although genetic diversity is decisive in 
favour of evolution in the environmental and historical variability,
the phenotypic variety can be ascribed to the randomness, intrinsic 
and extrinsic. Cells, being exposed to the diverse environment, 
must respond to the fluctuating external signal to control their fate 
\cite{Tsimring2014}.

A two-component system (TCS) is the most ubiquitous and the 
most compact signal transduction machinery observed in bacteria 
and in some plants and fungi \cite{Laub2007,Goulian2010,Groisman2016}. 
The TCS is a type of phosphotransfer cascade like the 
MAPK cascade, but differs in the molecular mechanism \cite{Stock2000}. 
In TCS, only one molecule of ATP is consumed and gets transferred 
through the cascade, whereas in the case of MAPK, at every 
phosphorylation step ATP is consumed. The TCS is composed 
mainly of two parts: A membrane-bound sensor kinase (SK) protein, 
and a cytoplasmic response regulator (RR) protein. The RR protein, 
activated by the phosphorylation, acts as a transcription factor for 
downstream genes or facilitates another target like flagellar locomotive 
switch \cite{Micali2016}. The simplest and modular TCS, which are 
efficiently present in the broad spectrum of efficacy, overtake popularity 
than other signal transduction motifs due to its inherent plasticity in 
noise regulation. Earlier studies on functionality of the SK 
\cite{Maity2014,Ortega2002,Yang2015}, input-output robustness of the 
phosphotransfer module \cite{Shinar2007, Batchelor2003}, bistability 
due to TCS mediated gene regulation 
\cite{Igoshin2008,Ghosh2011,Hoyle2012}, different stochastic switching 
responses \cite{Kierzek2010} and stochasticity induced active state 
locking and growth-rate dependent bistability \cite{Wei2014}
have been reported. 
Although most of the reported TCS have an autoregulatory feature,
the know-how of noisy gene regulation coupled with the TCS 
signaling pathway has not been explored till date. Within the purview 
of single-cell scenario, the effect of fluctuations can not be neglected 
in the TCS signaling pathway as fluctuations, whatever may be its 
source, has a key role in the gene regulation as well as in the 
post-translational modification that is taking place in the noisy cellular 
environment \cite{Sanchez2008}. In the present study, our prime 
interest is to quantify fluctuations in the TCS that has a feedback in 
the form of autoregulation at the operon of the two proteins, the sensor 
kinase and the response regulator. Depending on the nature of 
fluctuations of the extracellular signal, the signaling cascade must 
orient itself as to sense and respond appropriately. To this end, the 
phosphorylated response regulator ($R_P$) is considered as the 
output of the network. Hence, to study the propagation of noise 
through the cascade, we need to quantify the noisy characteristics 
of $R_P$ due to the external signal, the inducer ($I$). In this paper, 
we compute several experimentally realisable quantities, viz., variance, 
Fano factor, etc. We also calculate the information processing through 
the network as TCS transmits information of the alteration of the
environment.


\begin{figure}[!t]
\includegraphics[width=1.0\columnwidth,angle=0]{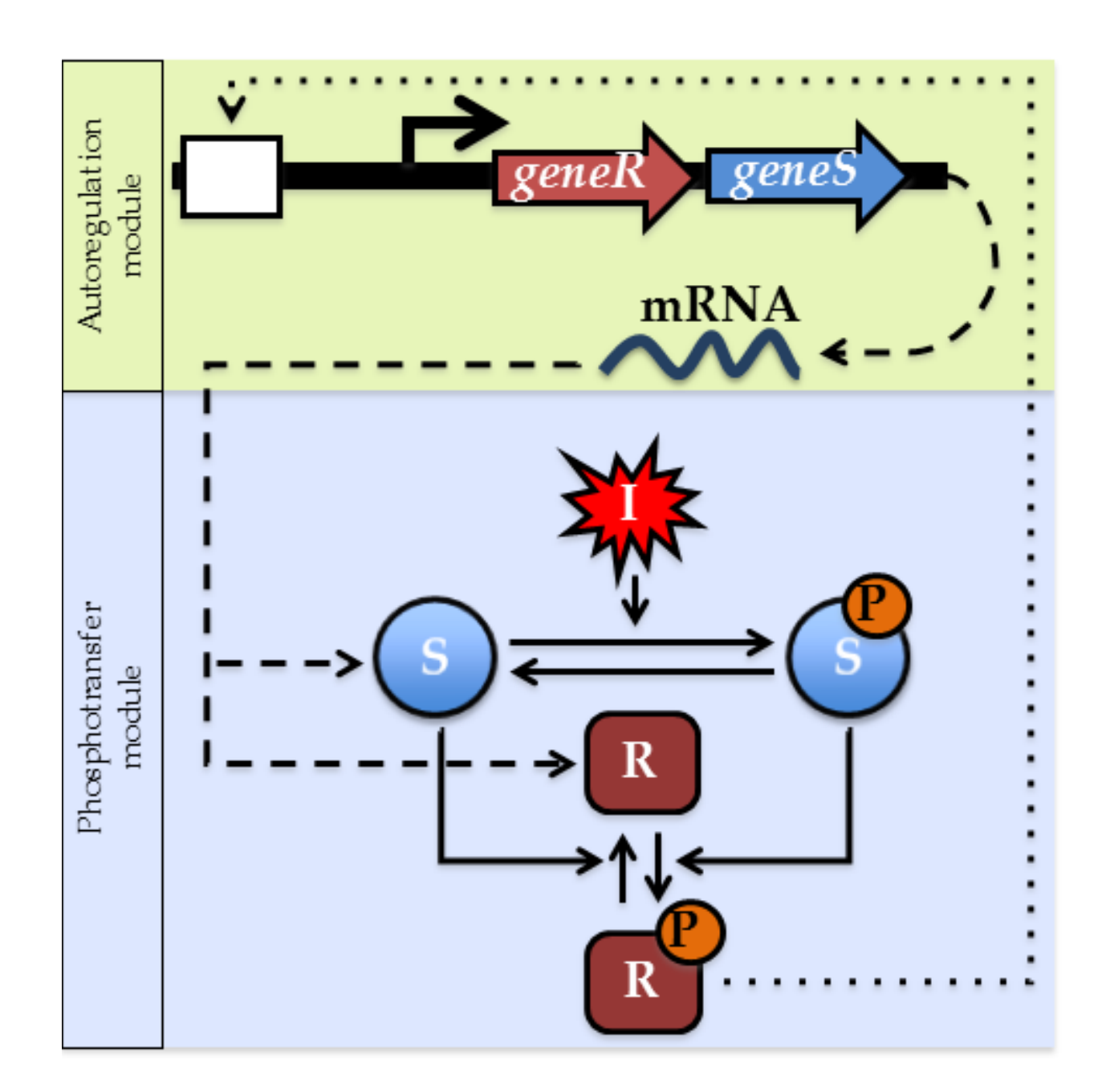}
\caption{(color online) The two-component system.
Schematic diagram of a generic two-component system composed of 
the phosphotransfer and the autoregulation module. The red asteroid
shows external inducer $I$. The blue sphere stands for the sensor kinase, 
$S$. Similarly, the brown box represents the response regulator, $R$. 
$P$ in orange sphere stands for the phosphate group. The wavy blue 
line is for the mRNA generated from the operon.
The solid lines with arrowhead stands for phosphotranfer kinetics.
The dashed lines takes care of both transcription and subsequent
translation. The dotted line represents the positive feedback on the
operon.
}
\label{fig1}
\end{figure}

In the last section of the paper, we present a population switch that
makes transition from bistable to monostable state and vice versa. 
Bistability is a widely observed 
phenomenon in biochemical networks with sufficient nonlinearity 
\cite{Tiwari2011}. In the case of deterministic systems, with weak 
nonlinearity, bistability can not be observed. However, if one goes for 
small reaction volume, the generated stochasticity may induce bistability 
\cite{Friedman2006,Bishop2010}. Nowadays with the increased use of 
the methods like fluorescence microscopy and flow cytometry, one may
decipher the noise induced heterogeneity in a clonal population 
\cite{Raj2008,Spiller2010}. The advantage of a microscopic stochastic 
study over the macroscopic deterministic observation is that the same 
mathematical construct of a TCS network shows monostable as well 
as bistable property depending on the system size.  Bistable switches 
do not have any clue of the genetic changes, hence they are epigenetic. 
Bistability has been observed in \textit{E. coli} by persister cells and in 
\textit{B. subtilis} through spore formation and cannibalism 
\cite{Dubnau2006}. But the mechanism is different for the two species. 
The first one happens due to positive autoregulation of the master gene. 
In the second one, there exist a pair of mutual repressors. In the 
current model, we are interested in the first type, positive autoregulation
driven bistability. To this end, we examine the TCS dynamics using a
stochastic approach for a range of parameter set to observe the transition 
from unimodal to a bimodal distribution.

\section{The Model}

The system, here we are dealing with, is a generic bacterial two-component 
signal transduction machinery that gets activated by an external stimulus.
While activated, it regulates one or several downstream genes. The full signal
transduction pathway may be divided into two interconnected modules, 
the autoregulation module and the phosphotransfer module. The autoregulation 
module is driven by the latter. In the phosphotransfer module, the external 
stimulus or the inducer $I$ triggers the autophosphorylation of the 
membrane-bound sensor kinase (SK) $S$ at the conserved Histidine residue 
to produce the phosphorylated form of $S$, $S_P$.  The phosphate group is 
then transferred to the Aspartate domain of the cognate cytoplasmic response
regulator (RR) $R$ via a phosphotransfer mechanism along with the formation 
of $R_P$. Once produced $R_P$ regulates the binding of RNA polymerase to 
the DNA of several downstream genes. In addition, it regulates the activity of 
its operon present in the autoregulation module (see Fig.~1). The 
phosphorylated RR becomes dephosphorylated by the unphosphorylated SK,
thus acting as a phosphatase. The ability of the SK to act as kinase as well as
phosphatase assigns it a bifunctional characteristic 
\cite{Laub2007,Tiwari2010}
and allows one to place it in a broad category of functional motifs, the 
paradoxical components \cite{Hart2013}. For a bifunctional TCS SK can thus 
act as a source and a sink of the phosphate pool. The relevant kinetics of the 
full network along with the kinetic rate parameters are tabulated in Table~1.
At this point it is important to mention that, in certain TCS, the phosphatase 
activity of the sensor kinase is absent thus making the TCS monofunctional 
\cite{Laub2007}.

To incorporate fluctuations due to different noise processes, we opt for 
Langevin method and calculate various physical entities. The extracellular 
noise comes from the environmental stimulus $I$. On the other hand, the 
gene regulation and the post-translational modification contribute to the 
intracellular noise. The Langevin equation of the inducer can be written as
\begin{equation}
\label{eq1}
\frac{dI}{dt} = k_{sI} - k_{dI} I + \xi_I ,
\end{equation}

\noindent with
$\langle \xi_I (t) \rangle = 0$, 
$\langle \xi_I (t) \xi_I (t+\tau) \rangle = (k_{sI}+k_{dI} \langle I \rangle) \delta (\tau)
= 2 k_{dI} \langle I \rangle \delta (\tau)$.
Here $\langle I \rangle$ is the mean (ensemble average) inducer level 
at steady state. $\xi_{I}$ is the extracellular fluctuations incorporated by 
the external stimulus. The phosphotransfer reaction kinetics shown in 
Table~I with the associated stochasticity can be expressed as
\begin{eqnarray}
\label{eq2}
\frac{dS_P}{dt} &=& k_{ap} I (S_T - S_P) - k_{adp} S_P - k_{k} S_P (R_T - R_P) 
\nonumber \\ 
&& - k_{dp} S_P + \xi_{S_P} ,\\
\label{eq3}
\frac{dR_P}{dt} &=& k_{k} S_P (R_T - R_P) - k_{p} R_P (S_T - S_P)  
\nonumber \\ 
&& - k_{dp} R_P + \xi_{R_P} ,
\end{eqnarray}

\noindent where $S_T$, $S_P$, $R_T$ and $R_P$ are the pool of total 
and phosphorylated form of $S$ and $R$ respectively. We note that while
writing Eqs.~(\ref{eq2}-\ref{eq3}) we consider simple bimolecular kinetics
instead of Michaelis-Menten kinetics. This is a valid approximation as long
as the Michaelis-Menten complex are low in concentration. Also,
consideration of bimolecular kinetics makes our subsequent analytical
calculation tractable. As a result,
$S (t) + S_P (t) = S_T (t)$ and $R (t) + R_P (t) = R_T (t)$, 
which we have used while writing Eqs.~(\ref{eq2}-\ref{eq3}). Here, $S_T(t)$ 
and $R_T(t)$ are not constant quantity, rather they keep on changing with 
time and only reach a constant value at steady state for a fixed value of
inducer level.
The $R_P$ mediated transcription and the expression kinetics of the two 
different pools of proteins (SK and RR) is formulated as follows
\begin{eqnarray}
\label{eq4}
\frac{dm}{dt}&=&k_{sm} f(R_P) - k_{dm}m + \xi_{m},\\
\label{eq5}
\frac{dS_T}{dt} &=& k_{ss} m - k_{dp} S_T + \xi_{S_T} ,\\
\label{eq6}
\frac{dR_T}{dt} &=& k_{sr} m - k_{dp} R_T + \xi_{R_T} .
\end{eqnarray}

\noindent Here $f(R_P) = R_P/(R_P+K)$, $K$ is the dissociation constant.
$\xi_m$ is the transcriptional noise associated with the synthesis and
degradation of mRNA, $m$. $\xi_{S_T}$ and $\xi_{R_T}$ are 
translational noise. The additive noise terms $\xi_{S_P}$, $\xi_{R_P}$, $\xi_{m}$, 
$\xi_{S_T}$ and $\xi_{R_T}$ take care of fluctuations in the copy number of 
$S_P$, $R_P$, $m$, $S_T$ and $R_T$, respectively. The noise terms are 
independent and Gaussian distributed with the statistical properties
\cite{Maity2014,Elf2003,Swain2004,Paulsson2004,Nicola2006,Warren2006,Kampen2007,Mehta2008,Maity2015,Grima2015,Biswas2016}
$\langle \xi_{S_P} (t) \rangle = \langle \xi_{R_P} (t) \rangle = 
\langle \xi_{m} (t) \rangle = \langle \xi_{S_T} (t) \rangle =
\langle \xi_{R_T}(t) \rangle$ $ = 0$ and
\begin{eqnarray*}
\langle \xi_{S_P} (t) \xi_{S_P} (t+\tau) \rangle & = & 2 k_{ap} 
[\langle S_T \rangle - \langle S_P \rangle] \langle I \rangle  \delta (\tau) , \\
\langle \xi_{R_P} (t) \xi_{R_P} (t+\tau) \rangle & = & 2 k_{k}  \langle S_P \rangle 
[\langle R_T \rangle - \langle R_P \rangle]  \delta (\tau), \\
\langle \xi_{m} (t) \xi_{m} (t+\tau) \rangle & = & 2 k_{dm} \langle m \rangle \delta (\tau) , \\
\langle \xi_{S_T} (t) \xi_{S_T} (t+\tau) \rangle & = & 2 k_{dp} \langle S_T \rangle \delta (\tau) , \\
\langle \xi_{R_T} (t) \xi_{R_T} (t+\tau) \rangle & = & 2 k_{dp} \langle R_T \rangle \delta (\tau).
\end{eqnarray*}

\noindent
The cross-correlation between $\xi_{S_P}$ and $\xi_{R_P}$ arises 
naturally due to kinetics shown by Eqs.~(\ref{eq2}-\ref{eq3})
\cite{Maity2014,Swain2004,Nicola2006},
\begin{eqnarray*}
\langle \xi_{S_P} (t) \xi_{R_P} (t+\tau) \rangle = - k_{k} \langle S_P \rangle
[\langle R_T \rangle - \langle R_P \rangle] \delta (\tau).
\end{eqnarray*}

\noindent 
Here, $\langle \cdots \rangle$ represents ensemble average evaluated 
at steady state. While writing the noise correlation in the present work 
we have used constant noise intensity evaluated at steady state, an 
approximation, that makes the following analytical calculation tractable.
Linearization of Eq.~(\ref{eq1}) and 
Eqs.~(\ref{eq2}-\ref{eq6}) around the mean value at steady state, i.e.,  
$I = \langle I \rangle + \delta I$, $S_P = \langle S_P \rangle + \delta S_P$, 
$R_P = \langle R_P \rangle + \delta R_P$, $m = \langle m \rangle + \delta m$, 
$S_T = \langle S_T \rangle + \delta S_T$ and 
$R_T = \langle R_T \rangle + \delta R_T$ yields
\begin{equation}
\label{eq8}
\frac{d}{dt} \mathbf{\delta X(t)} = \mathbf{J}_{X = \langle X \rangle} \mathbf{ \delta X (t)} 
+ \mathbf{\Xi (t)}.
\end{equation}

\noindent Here, $\mathbf{\delta X}$ is the fluctuations matrix. $\mathbf{\Xi}$ 
and $\mathbf{J}$ are the noise matrix and the Jacobian matrix of the averages 
evaluated at steady state, respectively (see Appendix for explicit form of the
matrices). To solve Eq.~(\ref{eq8}) we write the Lyapunov equation at steady 
state
\cite{Elf2003,Paulsson2004,Kampen2007,Keizer1987,Paulsson2005,Gardiner2009,Grima2011}
\begin{equation}
\label{eq9}
\mathbf{J} \mathbf{\sigma} + \mathbf{\sigma} \mathbf{J}^T + \mathbf{D} = 0,
\end{equation}

\noindent where $\mathbf{\sigma}$ is the covariance matrix and 
$\mathbf{D} = \langle \mathbf{\Xi \Xi}^T \rangle$ is the diffusion matrix
(see Appendix).
To quantify all the network properties and the information transmission, 
we solve the Lyapunov equation (\ref{eq9}) and evaluate the
elements of the covariance matrix $\mathbf{\sigma}$
\begin{eqnarray*}
{\bf \sigma} = \left (\begin{array}{cccccc}
\sigma^2_I		&	\sigma^2_{IS_P}	&	\sigma^2_{IR_P}		&	\sigma^2_{Im}		&	\sigma^2_{IS_T}		&	\sigma^2_{IR_T} \\
\sigma^2_{S_PI}	&	\sigma^2_{S_P}	&	\sigma^2_{S_PR_P}		&	\sigma^2_{S_Pm}	&	\sigma^2_{S_PS_T}		&	\sigma^2_{S_PR_T}\\
\sigma^2_{R_PI}	&	\sigma^2_{R_PS_P}	&	\sigma^2_{R_P}		&	\sigma^2_{R_Pm}	&	\sigma^2_{R_PS_T}		&	\sigma^2_{R_PR_T}\\
\sigma^2_{mI}		&	\sigma^2_{mS_P}	&	\sigma^2_{mR_P}		&	\sigma^2_{m}		&	\sigma^2_{mS_T}		&	\sigma^2_{mR_T}\\
\sigma^2_{S_TI}	&	\sigma^2_{S_TS_P}	&	\sigma^2_{S_TR_P}		&	\sigma^2_{S_Tm}	&	\sigma^2_{S_T}		&	\sigma^2_{S_TR_T}\\
\sigma^2_{R_TI}	&	\sigma^2_{R_TS_P}	&	\sigma^2_{R_TR_P}		&	\sigma^2_{R_Tm}	&	\sigma^2_{R_TS_T}		&	\sigma^2_{R_T}
\end{array} 
\right )
\end{eqnarray*}

\noindent
The elements of $\mathbf{\sigma}$ are the variance and covariances of 
the subscripted quantities.

For numerical validation of the results calculated analytically, we adopt 
stochastic simulation algorithm (SSA) \cite{Gillespie1976,Gillespie1977}.
The kinetics and propensity associated with each chemical reaction are 
tabulated in Table~1. The simulation results presented in the following 
section are average of $10^6$ independent trajectories. During simulation, 
each independent run starts with a constant pool of $S$ and $R$ 
that takes care of basal level of the two proteins. In the absence of the stimulus 
$I$ ($k_s = 0$), only the degradation of $S$ and $R$ are operative along
with the constant pool of the same. In such situation, the concentration
of proteins goes down to zero at steady state. Once the inducer assumes 
a non-zero value ($k_s \neq 0$) it activates the phosphotransfer cascade 
and generates the pool of the transcription factor $R_P$, thus providing a 
positive feedback on the operon. As a result, the full signaling network 
(autoregulation+phosphotransfer) becomes operative.

\section{Results and Discussions}

The utility of the TCS is to respond rapidly by properly transducing 
an external signal. The response, here we are dealing with, is the 
transcriptional autoregulation at the operon of SK and RR along with 
other downstream genes by the phosphorylated response regulator 
$R_P$. The pool of $R_P$ in turn gets accumulated by the inducer 
mediated autophosphorylation and subsequent phosphotransfer of 
the sensor kinase $S_P$. The stochastic fluctuations can be calculated 
through different statistical measures for the current model. As $R_P$ 
is considered as the output of the full TCS, the common measures of 
fluctuations, like variance, Fano factor are observed for $R_P$. The 
fidelity or the reliability of the present signaling cascade can be also 
measured in terms of mutual information, network gain and intrinsic 
noise. From the biological outlook, the sources of different noise 
processes are widespread inside and outside the cell. Some sources 
of randomness are shared by different proteins, correlated through 
interaction in the post-translational modification or due to sharing common 
expression operon. If we observe the post-translational reactions like 
the phosphotransfer between $S$ and $R$, the source of the fluctuations 
may differ from  that of the gene regulation. Here, to quantify all the 
network properties, we propose the inducer, $I$ to be the input and 
$R_P$ to be the network output as $R_P$ being of prime interest 
since it regulates many of the downstream genes. The classification 
of the source of noise is difficult for such a large and modular network. 
But one can quantify the fluctuations through the aforesaid different 
measures.

The entire observation in this paper is partitioned into four segments. 
First, we focus on the phosphotransfer module to quantify fluctuations 
over a range of the external signal $I$. Second, we concentrate on the 
stochastic transcriptional process, which shows the dynamics of mRNA 
fluctuations in response to the signal mediated feedback of the response 
regulator. Third, we focus on the entire network to decipher different 
stochastic network properties, which reflect the reliability of the network 
performance. The final segment deals with the phosphorylated response 
regulator distribution at steady state to find the role of the switch in 
characterising the stability at the steady state.

 
\begin{figure}[!t]
\includegraphics[width=0.8\columnwidth,angle=-90]{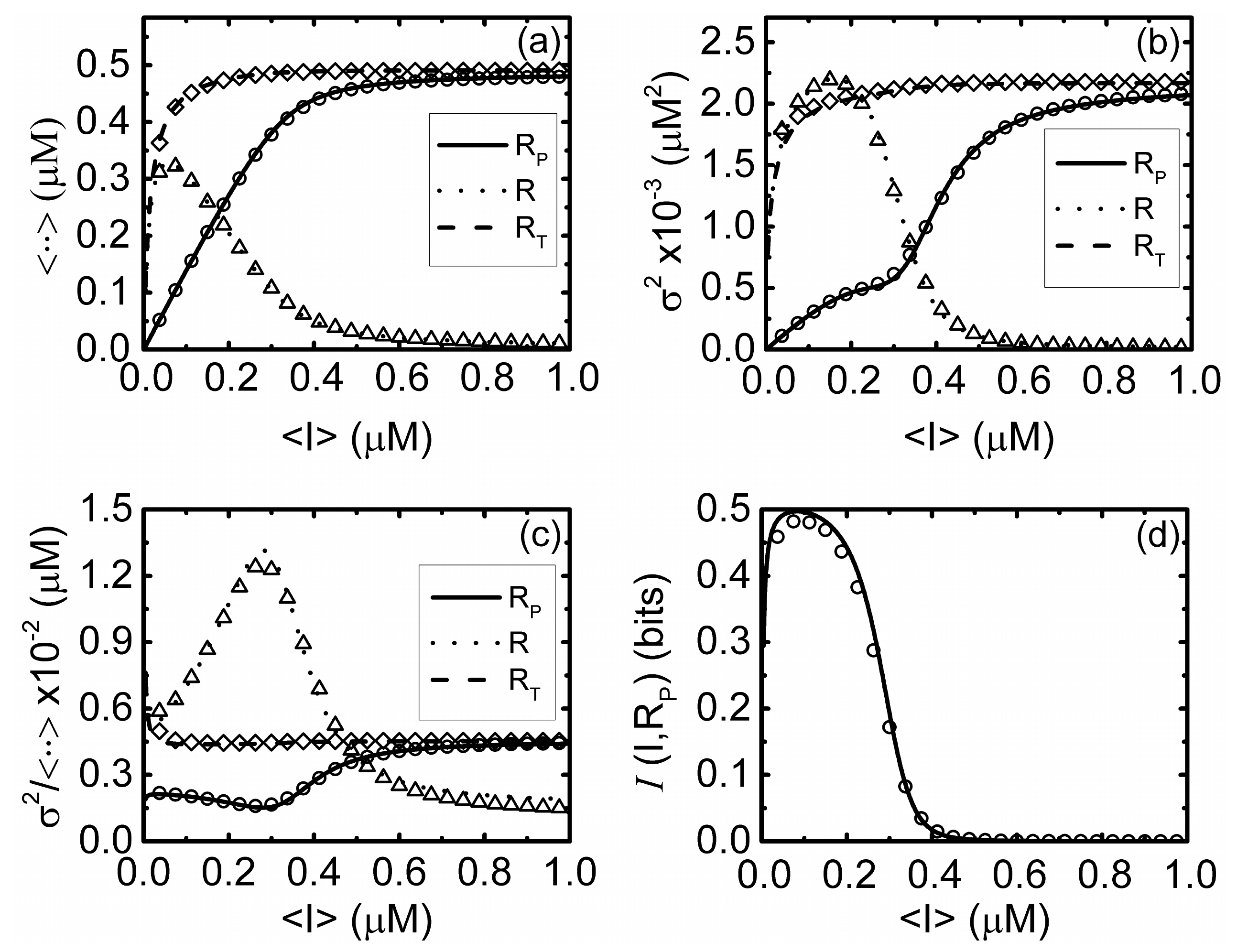}
\caption{Fluctuations in response regulator.
Different statistical measures for the pool of response regulator. The lines 
indicate the theoretical values of $R_P$ (solid), $R$ (dotted) and $R_T$
(dashed) and the symbols stand for the numerical values of $R_P$ (circle), 
$R$ (triangle) and $R_T$ (diamond). (a) averages 
($\langle R_P \rangle, \langle R \rangle, \langle R_T \rangle$), 
(b) variances ($\sigma^2_{R_P}, \sigma^2_R, \sigma^2_{R_T}$)
and (c) Fano factors ($\sigma^2_{R_P}/\langle R_P \rangle, 
\sigma^2_R/\langle R \rangle, \sigma^2_{R_T}/\langle R_T \rangle$) 
are plotted as a function of average signal strength $\langle I \rangle$. 
The mutual information ${\cal I} (I,R_P)$ between the the input signal $I$ 
and the output $R_P$ is shown in (d). The symbols are generated using 
stochastic simulation algorithm \cite{Gillespie1976,Gillespie1977} and
the lines are due to theoretical expressions.
}
\label{fig2}
\end{figure}

\subsection{Fluctuations in the protein pool}

While sensing the external stimulus $I$, $S$ gets itself phosphorylated 
and triggers the flow of signal. During phosphotransfer to $R$ through 
the kinase activity, $S_P$ transmits the noisy signal to the pool of $R_P$.
In the opposite, when $S$ removes the phosphate from $R_P$ due to 
its phosphatase activity, there is no clue of noise reduction. The pool of 
total response regulator $R_T$ remains partly unphosphorylated ($R$) 
at low signal strength (Fig.~2(a)). At this level, the variance of $R_P$ rises 
sharply in contrast to the quick fall of the variance of $R$. The net 
fluctuations level in $R_T$ remains consistent with the model assumptions,
\begin{eqnarray}
\label{eq10}
\langle R_T\rangle &=& \langle R_P \rangle + \langle R \rangle,\\
\label{eq11}
\sigma^2_{R_T} &=& \sigma^2_{R_P} + \sigma^2_{R} + 2\sigma^2_{R_PR},\\
\label{eq12}
\sigma^2_{R} &=& \sigma^2_{R_P} + \sigma^2_{R_T} - 2\sigma^2_{R_PR_T},
\end{eqnarray}

\noindent
where $\langle \cdots \rangle$ stands for the expectations of the protein 
pool and $\sigma^2$ for the variances and covariances. As the signal rises, 
the pool of the response regulator gets fully phosphorylated and the variance 
of $R_P$, $\sigma^2_{R_P}$ gets saturated with $\sigma^2_{R_T}$ (Fig.~2(b)). 
The diminished fluctuations in the pool of $R$, $\sigma^2_{R}$ is associated 
with the decaying pool of $R$. Also, we focus on the Fano factor 
($\sigma^2/\langle \cdots \rangle$), which implies the amount of fluctuations 
for unity of molecules (Fig.~2(c)). The Fano factor of $R_T$ remains quite 
invariant over the range of signal establishing that switching of the response 
regulator in two phases ($R$ and $R_P$) does not incorporate additional 
fluctuations in the molecular pool of $R_T$.


\begin{figure}[!t]
\includegraphics[width=0.8\columnwidth,angle=-90]{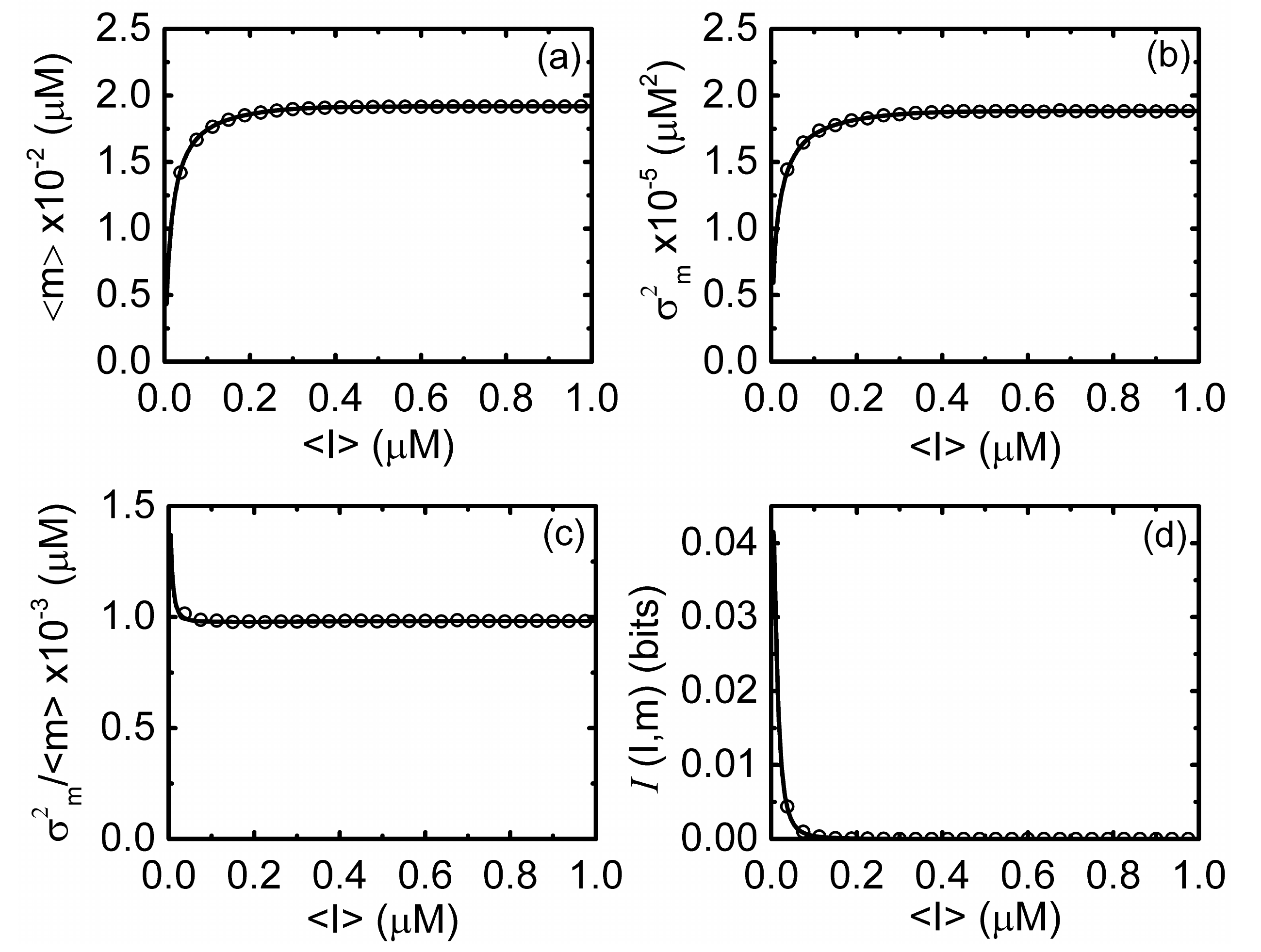}
\caption{Fluctuations in mRNA.
The (a) mean ($\langle m \rangle$), (b) variance ($\sigma^2_m$), 
and (c) Fano factor ($\sigma^2_m/\langle m \rangle$) are plotted with 
respect to the average signal strength $\langle I \rangle$. In (d) the 
transmission of information (${\cal I} (I,m)$) in the autoregulation
mediated transcription is shown. The symbols are generated using 
stochastic simulation algorithm \cite{Gillespie1976,Gillespie1977} 
and the lines are due to theoretical expressions.
}
\label{fig3}
\end{figure}

\subsection{Transcriptional randomness}

The central dogma of the molecular biology involves two steps - gene 
transcription, to produce mRNA and further in turn translation, to protein. 
In the present work, we are interested in the zeroth-order $R_P$-regulated 
transcription from the operon of $R$ and $S$. The accumulation of the 
mRNA pool is regulated by the first-order degradation of mRNA. $R_P$, 
activated from the phosphotransfer module, binds at the promoter in the
form of a feedback with Hill function kinetics, $f(R_P) = R^H_P/(K^H+R^H_P)$. 
Here, $K$ (= $20$ $\mu$M) is the equilibrium dissociation constant of the 
transcription factor $R_P$ to the promoter. The Hill coefficient $H=1$, as 
considered in the current model, corresponds to a positive feedback without 
any cooperativity. To monitor the stochasticity introduced in the transcriptional 
kinetics, we quantitate the fluctuations in the pool of the mRNA. Fig.~3(a-c) 
shows the statistical properties of mRNA expression in response to the signal. 
For a wide
range of signal the variance of mRNA, $\sigma^2_m$ remains constant and 
is equal to the average, $\langle m \rangle$ which makes Fano factor equals 
to unity, a signature of Poissonian statistics. We note that the equality relation 
between the variance and the mean value can be checked when both the
quantities are expressed in copy numbers/cellular volume instead of 
concentration using the relation 1M $\approx 1.024 \times 10^9$ copy 
number/cellular volume. The overall statistical features observed in the
mRNA dynamics suggests that the feedback controlled autoregulation 
and subsequent transcription behave like a fluctuational buffer in the whole 
network. As a result, a weak information transmission occurs through the 
autoregulation module (Fig.~3(d)).


\begin{figure}[!t]
\includegraphics[width=0.8\columnwidth,angle=-90]{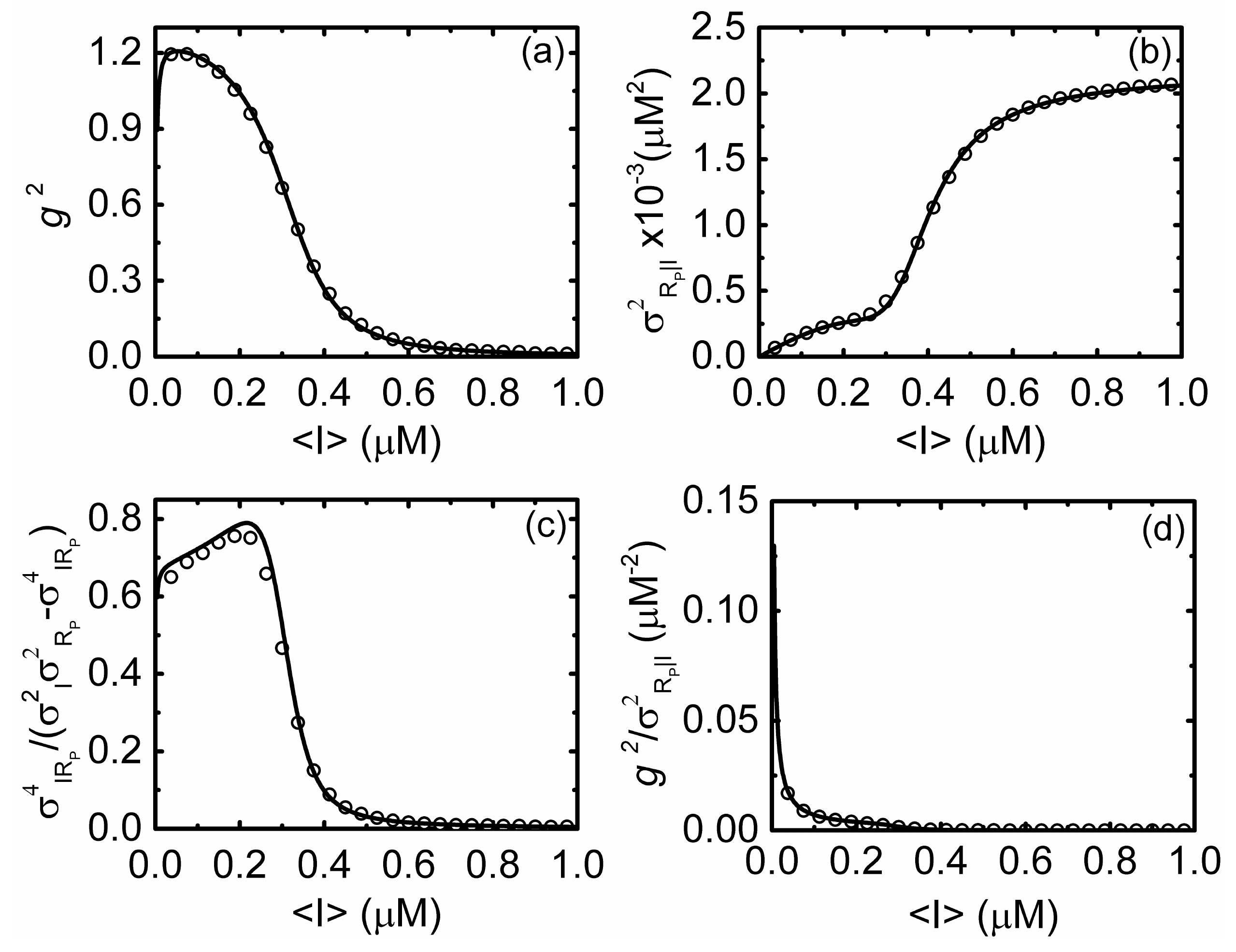}
\caption{The signaling performance of the TCS.
The (a) network gain ($g^2$), (b) intrinsic noise ($\sigma^2_{R_P|I}$), 
(c) signal-to-noise ratio, and (d) gain-to-noise ratio are presented 
for the full range of the signal $\langle I \rangle$ at steady state.
The symbols are generated using stochastic simulation algorithm 
\cite{Gillespie1976,Gillespie1977} and the lines are due to theoretical 
expressions.
}
\label{fig4}
\end{figure}

\subsection{The TCS signaling network}

In this subsection, we discuss the motivation for considering the TCS 
motif as a signal transducer. To this end, it is crucial to quantify both 
the gain and the noise, not in isolation of them. To elucidate the network 
performance, we consider four different measures: network gain ($g^2$), 
intrinsic noise ($\sigma^2_{R_P|I}$), signal-to-noise ratio (SNR), and 
gain-to-noise ratio (GNR). The quantitative measures can be defined 
\cite{Tostevin2010} as follows, as the TCS cascade satisfies the spectral 
addition rule \cite{Warren2006},
\begin{eqnarray}
\label{eq13}
g^2 &=& \frac{\sigma^2_{R_PI}}{\sigma^2_I},\\
\label{eq14}
\sigma^2_{R_P|I} &=& \sigma^2_{R_P} - g^2 \sigma^2_I,\\
\label{eq15}
SNR &=& \frac{\sigma^4_{IR_P}}{\sigma_I^2 \sigma_{R_P}^2 - \sigma_{IR_P}^4}
=\frac{g^2}{\sigma^2_{R_P|I}}\sigma^2_I,\\
\label{eq16}
GNR &=& \frac{g^2}{\sigma^2_{R_P|I}}.
\end{eqnarray}

\noindent
The gain, defined here, is different 
from the macroscopic gain \cite{Savageau1976,Goldbeter1981,Koshland1982}
which characterises the fold increment in 
the observable output for a constant signal. The stochastic network 
gain is a qualitative measure about the uncertainty estimation of the 
network input when the uncertainty in the output is well known. The 
GNR gives a better insight on this quantity. It is important to note that 
the mutual information has an estimation from the reciprocal of the 
GNR. In Fig.~4(a) the network gain for the current model is being 
illustrated with respect to the level of the inducer. At low signal 
strength, the phosphotransfer reactions show determining role in the 
network, where the fluctuations level of both $R$ and $R_P$ grow. 
Such growth implies the association of the randomness between 
inducer $I$ and both $R_P$ and $R$ increases. This association 
assists the elevation in the gain value. As the signal increases, the 
pool of the response regulator gets fully phosphorylated, and the 
fluctuations strength in the $R_T$ pool becomes constant. As a result, 
the gain and the GNR deplete (Fig.~4(a),4(d)).

The intrinsic noise, explains the randomness within the system, 
usually occurs through the inherently probabilistic biochemical 
reactions. Following the spectral addition rule fluctuations due to 
inducer ($\delta I$) has been eliminated from the fluctuations of 
the network output ($\delta R_P$) to purify the intrinsic noise of 
the TCS cascade (Eq.~(\ref{eq14})). Fig.~4(b) shows the intrinsic 
noise profile as a function of inducer level. The intrinsic noise is 
nearly equal to $\sigma^2_{R_P}$ (shown in Fig.~2(b)) as the 
contribution of $g^2$ is minuscule for increasing signal strength.

The SNR is a good measure of the fidelity for a signaling network. 
Better the SNR, the better is the reliability of the cascade. The 
mutual information is the quotient of reliability, and it depends on 
the SNR. In Fig.~4(c), we show the characteristic profile of SNR vs. 
inducer. The notion of this plot is to decrypt the effective signal 
transmission through the signaling channel. The SNR initially starts 
from a moderate value at a very low signal strength and maximises 
where the $R_P$ population reaches the maximum value. As a 
consequence of the sharp rise of the fluctuations in $R_P$ as well 
as the intrinsic noise, the SNR experiences a quick decay. From the 
inclination of fidelity, it is to infer that the TCS shows maximum reliability 
at low to moderate level of the signal $\langle I \rangle$ when the 
$R$ to $R_P$ transition rate is high.

\begin{widetext}


\begin{figure}[!t]
\includegraphics[width=0.4\columnwidth,angle=-90]{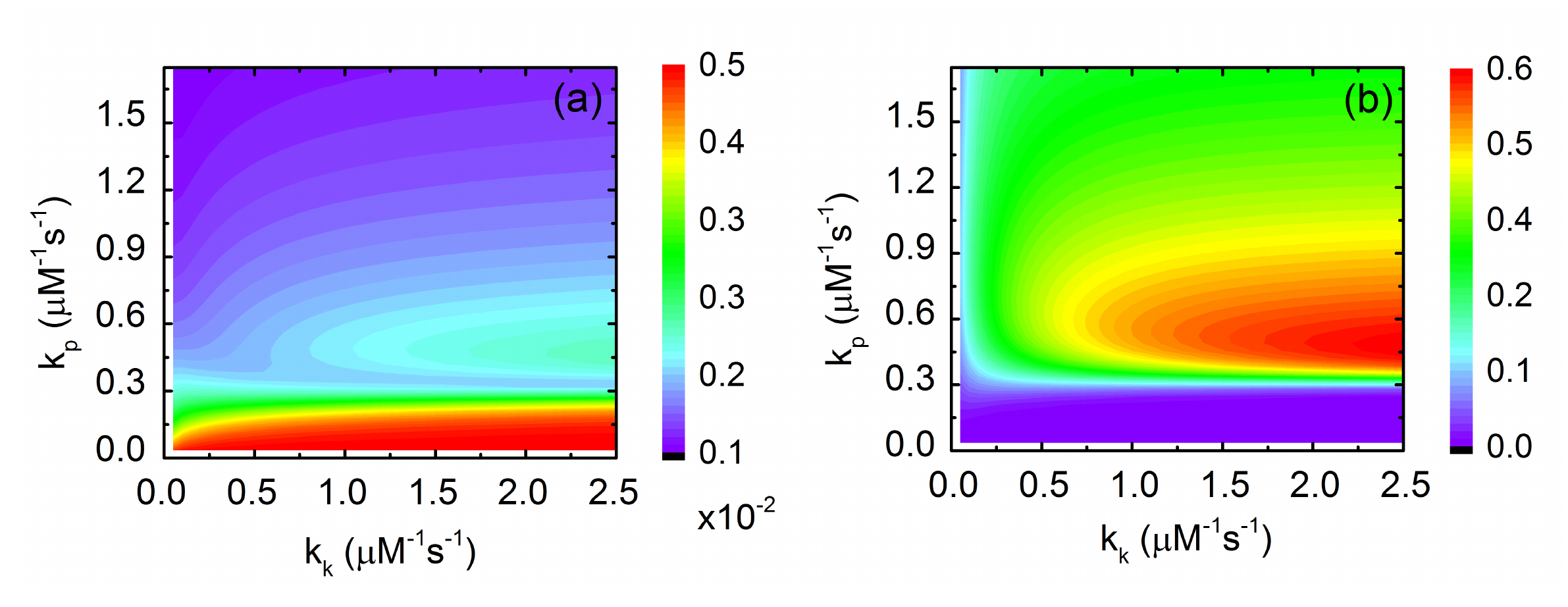}
\caption{(color online) The role of kinase and phosphatse rate.
The two dimensional map of (a) the Fano factor 
($\sigma^2_{R_P}/\langle R_P \rangle$) and (b) the mutual information 
(${\cal I}(I, R_P)$) as a function of the kinase and phosphatase rate 
constants of the phosphotransfer module. The maps are drawn 
using $\langle I \rangle =$ 0.15 $\mu$M. All the other rate parameters 
are according to Table~I.
}
\label{fig5}
\end{figure}

\end{widetext}

\subsection{Information transmission}

The functionality of a two-component system is to sense and respond 
appropriately to any extracellular signal. As we are considering the 
associated fluctuations in the signaling pathway, the quantitative 
reliability in terms of mutual information is to be measured. For a 
Gaussian system, the mutual information provides the channel 
capacity \cite{Shannon1948,Cover1991}. Considering the inducer 
($I$) as network input and phosphorylated response regulator 
($R_P$) as the output, one can quantify the mutual information
${\cal I}(I,R_P)$ between $I$ and $R_P$. The interplay between the 
signal and the network response can be verified when the association 
of the input-output fluctuations space is determined. According to the 
definition of Shannon \cite{Shannon1948}, for Gaussian noise processes,
the mutual information can be written as
\begin{equation}
\label{eq17}
{\cal I}(I,R_P) = \frac{1}{2} \log_2 
\left( 
1+ \frac{\sigma^4_{IR_P}}{\sigma_I^2 \sigma_{R_P}^2 - \sigma_{IR_P}^4} 
\right),
\end{equation}

\noindent
where the quantity 
$\sigma^4_{IR_P}/(\sigma_I^2 \sigma_{R_P}^2 - \sigma_{IR_P}^4)$ 
measures the fidelity 
(SNR). In Fig.~2(d), we show mutual information as a function of the 
inducer level. The information processing by the TCS shows a sharp 
growth followed by a decaying nature. Beyond a certain inducer level, 
the mutual information profile goes down implying attainment of the 
steady state of the response regulator pool.

Since the phosphotransfer module plays the dominant role in the TCS 
cascade, the investigation for the two critical reactions: kinase and 
phosphatase within the frame of mutual information is crucial. In Fig.~5, 
we show the contour map of Fano factor and mutual information
as a function of the phosphotransfer rate parameters: kinase, $k_k$ 
and phosphatase, $k_p$. While the Fano factor of $R_P$ measures 
the output fluctuations at the molecular level, ${\cal I}(I,R_P)$ quantitates 
the association between the input and output fluctuations level. So a 
comparative optimisation of the rate constants is executed here in the 
two maps. In Fig.~5(a), one can see that at very low phosphatase rate 
the fluctuations level in $R_P$ is maximum; else it is quite low. In the 
same regime, the information processing is very weak (Fig.~5(b)). The 
information processing maximises at a moderate phosphatase and high 
kinase level where the fluctuations level is remarkably low. This happens 
due to the low amount of $R_P$ at high phosphatase activity of the sensor 
protein, $S$.

For the autoregulated transcription module, the mutual 
information of the noisy gene expression can be written as
\begin{equation}
\label{eq18}
{\cal I}(I,m) = \frac{1}{2} \log_2 
\left( 
1+ \frac{\sigma^4_{Im}}{\sigma_I^2 \sigma_m^2 - \sigma_{Im}^4} 
\right).
\end{equation}

\noindent
The fluctuations space associated with mRNA shown in the profiles 
of variance and Fano factor (see Figs.~3(b,c)) is so narrow that the 
intersection of that with the fluctuations space of the inducer is minuscule. 
It gets also reflected in the profile of the mutual information ${\cal I}(I,m)$ 
in Fig.~3(d). With the enhancement of the signal strength, the information 
flow through the transcription reaction attenuates sharply. Since the gene 
regulation and the transcription in this network play the role of a noise buffer, 
the sharp decrease in the information processing capacity is well justified.


\begin{figure*}[!h]
\includegraphics[width=2.0\columnwidth,angle=0]{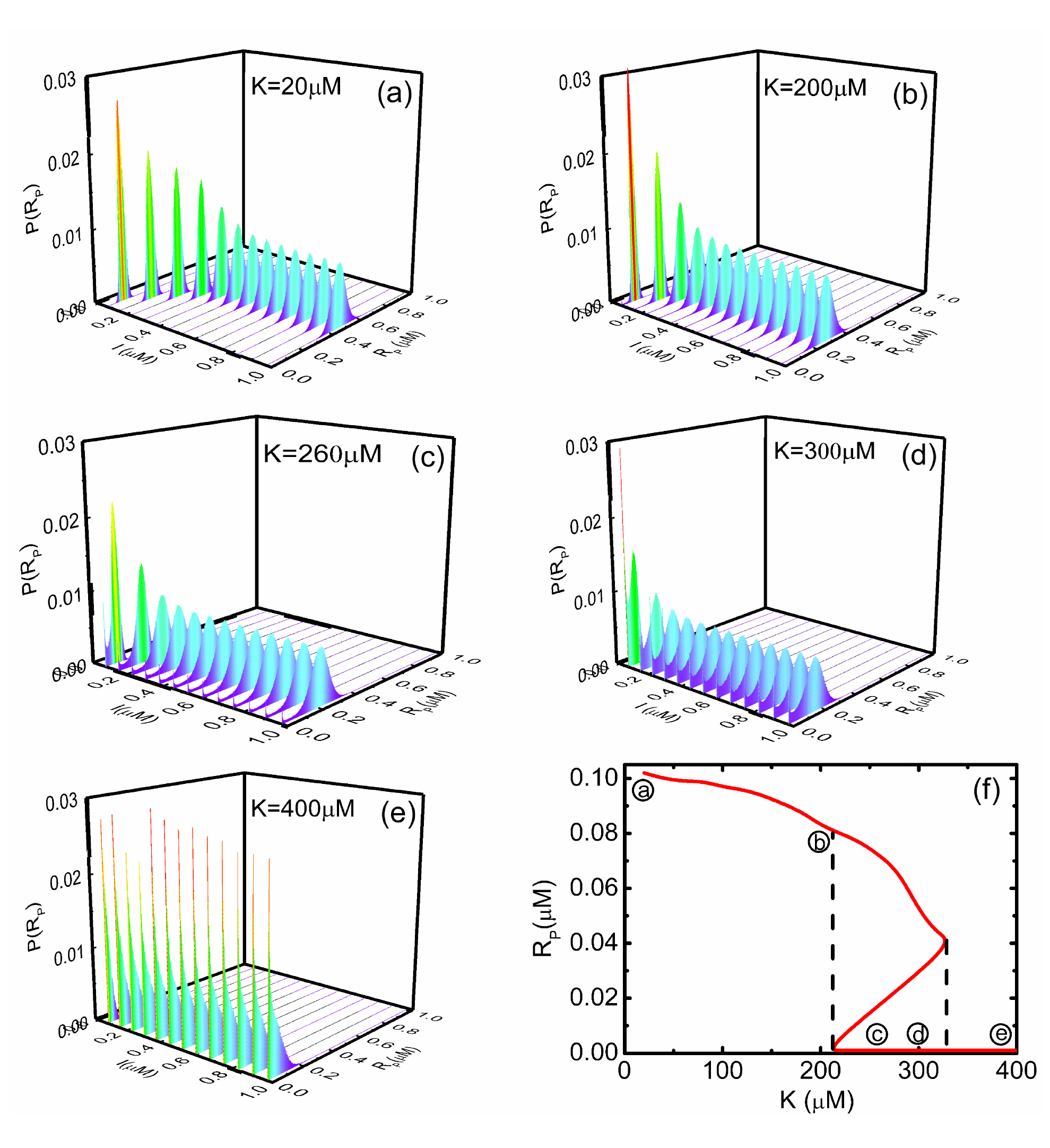}
\caption{(color online) Noise induced bimodality.
(a)-(e) The three dimensional histogram of the probability density 
of $R_P$ at steady state is plotted for different values of the 
dissociation constant $K$. The $x$-axis shows the level of signal, 
$I$ and the population of $R_P$ is given in the $y$-axis. (f) The 
mean population of $R_P$ with respect to the dissociation constant 
$K$ at low value of signal ($I=0.075$ $\mu$M). The region inside 
the dotted lines shows bimodality.
}
\label{fig6}
\end{figure*}

\subsection{Protein distribution and the stability switch}

In this section, we propose a numerical framework which explores 
the distribution of the phosphorylated response regulator protein at the 
steady state. The analysis of the two-component system ascribed 
above is within the purview of linear noise approximation.
Now, we want to extend the same network beyond the linear limit 
and examine some key features from the distribution profile. Here, 
we observe the noise propagation through the autoregulation module
under the control of the dissociation constant $K$ of the transcription 
factor $R_P$ at the promoter site, as different ranges of $K$ shows 
different probability distributions for a particular signal strength 
(Fig.~6(a-e)).

As the dissociation constant $K$ is the ratio of the two rate constants: 
The unbinding ($k_u$) and the binding ($k_b$) rate of the transcription 
factor, the variation in $K$ drives the full regulated expression through 
the alteration in the promoter kinetics. At lower $K$-value, binding of 
$R_P$ to the promoter is more stable, and the duration of the `ON' 
state of the promoter is much larger than the `OFF' state. Hence, the 
transcription reaction incorporates lesser fluctuations in further 
proceedings. 
With the increment of $K$ the time scales of the `ON' and the `OFF' 
state of the promoter appears to be comparable. As a result, two 
separate regime appears in the transcription timing, and this intrinsic 
stochasticity introduces distinct bimodality in the population of $R_P$. 
The presence of the weak nonlinearity in $f(R_P) = R_P/(K+R_P)$ 
for positive feedback shows a zero-order ultrasensitivity, which always 
produces single-valued steady-state concentration, i.e. monostability, 
when characterised deterministically without noise. But the numerical
simulation of the stochastic model explores the stability switch which 
produces multi-valued steady state, i.e., bimodality. In Fig.~6(f), one 
can see that the bimodal population of $R_P$ exists within a very 
narrow range of $K$. At the exterior of this range, the distribution is 
unimodal at the low or the high population. If we denote the small 
copy number state as the persisters, that frequently occurs in a 
bacterial population, the sharp switch between the two states must 
have an evolutionary and survival fate.

\section{Conclusions}

The present study reports a theoretical analysis of signal transduction
in bacterial two-component signalling pathway using a stochastic 
framework. The theoretical study makes a detailed discussion of the 
fluctuations propagation in the presence of an external signal. We
find that the switching of the response regulator in its two states
cannot produce more randomness in the protein pool but the
observed fluctuations appear to be conveyed through the cascade.
The observed Poissonian feature in mRNA pool predicts the
transcription system with weak bursts and less noise generation.
Our analysis also checks the reliability of the network in information 
processing. To this end, the fidelity of the network is examined 
through the signal-to-noise ratio. Our results suggest that the 
complete network has high efficiency at moderately low level of 
the signal. In this context, the modular structure of the network is 
being analysed. The phosphotransfer module takes the leading role 
in the propagation of fluctuations and information compared to the 
autoregulation module. The traditional metrics like variance and 
Fano factor measures the noise strength in the network but is 
unable to guess how fluctuations affect the process of information 
transmission. To take care of this issue, we focus on the signalling 
fidelity. Given the network gain and the intrinsic noise as two 
antagonistic metrics, an efficient network always tries to maximise 
the gain at the minimal inherent randomness. A multi-objective 
trade-off between these two different parameters makes a network 
more robust against the external stimulus.

The recent advent of experimental techniques facilitates the measure
of cellular and molecular fluctuations along with the information flow.
The TCS, a minimal example of the bacterial communication system, 
harvests reliable information transmission while making a cellular 
decision. The regulation of gene expression is the most crucial 
physiological function to be decided. But in the case of complex 
interconnected genetic networks, it is hard to decipher the exact 
path of noise incorporation or information flow. The modular approach 
to systems biology at single cell level may help in addressing this issue. 
The investigation of single-cell behaviour may also reveal the noise-driven 
bifurcation of a monoclonal population and the existence of a sharp 
switch between the phenotypes. This switching mechanism may 
evolve to adapt in an adverse environment. Rather than an evolutionary 
pressure, the noise makes the regulated gene expression develop 
towards an optimal state.

\begin{acknowledgments}
Thanks are due to Ayan Biswas and Alok Kumar Maity for fruitful discussion.
The authors acknowledge financial support from CSIR, India [01(2771)/14/EMR-II].
SKB is thankful to Bose Institute, Kolkata, India for a research fund through
Institutional Programme VI - Systems Biology.
\end{acknowledgments}

\section*{Data Availability}
All relevant data are within the paper.

\section*{Competing Interest} 
We have no competing interests.

\section*{Author Contributions}
TM SC SKB conceived and designed the study,
TM carried out the theoretical calculation and numerical simulation,
TM SKB analyzed the data,
TM SC SKB wrote the paper.
All authors gave final approval for publication.

\appendix*

\section{}

The fluctuations matrix $\delta \mathbf{X}$, the noise matrix $\mathbf{\Xi}$
and the Jacobian matrix $\mathbf{J}$ written in Eq.~(\ref{eq8}) are defined as,
\begin{eqnarray*}
\delta \mathbf{X} =
\left (
\begin{array}{c}
\delta I \\
\delta S_P \\
\delta R_P \\
\delta m \\
\delta S_T \\
\delta R_T
\end{array}
\right ),
\mathbf{\Xi} =
\left (
\begin{array}{c}
\xi_{I} \\
\xi_{S_P} \\
\xi_{R_P} \\
\xi_{m} \\
\xi_{S_T} \\
\xi_{R_T}
\end{array}
\right ), 
\end{eqnarray*}
\begin{eqnarray*}
\mathbf{J} = 
\left (
\begin{array}{cccccc}
       J_{I I}         &   J_{I S_P}              & J_{I R_P}              & J_{I m}             & J_{I S_T}              & J_{I R_T} \\
  J_{S_P I}     &   J_{S_P S_P}         & J_{S_P R_P}         & J_{S_P m}         & J_{S_P S_T}         & J_{S_P R_T}  \\
  J_{R_P I}      & J_{R_P S_P}        &  J_{R_P R_P}        &  J_{R_P m}       &  J_{R_P S_T}        &  J_{R_P R_T}  \\
      J_{m I}          & J_{m S_P}           &  J_{m R_P}            &  J_{m m}          &  J_{m S_T}            &  J_{m R_T}  \\
  J_{S_T I}       & J_{S_T S_P}        &  J_{S_T R_P}        &  J_{S_T m}        &  J_{S_T S_T}        &  J_{S_T R_T}  \\
  J_{R_T I}      & J_{R_T S_P}        &  J_{R_T R_P}        &  J_{R_T m}       &  J_{R_T S_T}        &  J_{R_T R_T}  
\end{array}
\right
) 
\end{eqnarray*}

\noindent with
\begin{eqnarray*}
&& J_{II}=-k_{dI}, J_{I S_P}=J_{I R_P}=J_{I m}=J_{I S_T}=J_{I R_T}= 0, \\
&& J_{S_P I}=k_{ap} (\langle S_T \rangle - \langle S_P \rangle), \\
&& J_{S_P S_P}=-k_{ap} \langle I \rangle - k_{adp} - k_{k} (\langle R_T \rangle - \langle R_P \rangle)- k_{dp} ,\\
&& J_{S_P R_P}=k_{k} \langle S_P \rangle, J_{S_P m}=0, J_{S_P S_T}=k_{ap} \langle I \rangle, \\
&& J_{S_P R_T}=-k_{k} \langle S_P \rangle, J_{R_P I}=0,\\
&& J_{R_P S_P}=k_{k} (\langle R_T \rangle - \langle R_P \rangle) + k_{p} \langle R_P \rangle, \\
&& J_{R_P R_P}=-k_{k} \langle S_P \rangle -k_{4} (\langle S_T \rangle - \langle S_P \rangle)-k_{dp}, J_{R_P m}=0, \\
&& J_{R_P S_T}=-k_{p} \langle R_P \rangle, J_{R_P R_T}=k_{k} \langle S_P \rangle, J_{m I}=0, J_{m S_P}=0, \\
&& J_{m R_P}=k_{sm}\left[\frac{\delta f(R_P)}{\delta R_P}\right]_{s.s}=k_{sm}\frac{K}{(\langle R_P\rangle+K)^2}, \\
&& J_{m S_T}= J_{m R_T}= J_{S_T I}= J_{S_T S_P}= J_{S_T R_P}=0, \\
&&  J_{S_T m}=k_{ss}, J_{S_T S_T}=-k_{dp}, J_{m m}=-k_{dm}, \\
&& J_{S_T R_T}=J_{R_T I}= J_{R_T S_P}=J_{R_T R_P}=0, \\
&& J_{R_T m}=k_{sr}, J_{R_T S_T}=0, J_{R_T R_T}=k_{dp}.
\end{eqnarray*}

The explicit form of the diffusion matrix written in Eq.~(\ref{eq9}) is
\begin{eqnarray*}
{\bf D} = \left (\begin{array}{cccccc}
	D_{I I}		&	D_{I S_P}		&	D_{I R_P}		&	D_{I m}		&	D_{I S_T}		&	D_{I R_T} \\
	D_{S_P I}		&	D_{S_P S_P}	&	D_{S_P R_P}	&	D_{S_P m}	&	D_{S_P S_T}	&	D_{S_P R_T} \\
	D_{R_P I}		&	D_{R_P S_P}	&	D_{R_P R_P}	&	D_{R_P m}	&	D_{R_P S_T}	&	D_{R_P R_T} \\
	D_{m I}		&	D_{m S_P}	&	D_{m R_P}	&	D_{m m}		&	D_{m S_T}	&	D_{m R_T} \\
	D_{S_T I}		&	D_{S_T S_P}	&	D_{S_T R_P}	&	D_{S_T m}	&	D_{S_T S_T}	&	D_{S_T R_T} \\
	D_{R_T I}		&	D_{R_T S_P}	&	D_{R_T R_P}	&	D_{R_T m}	&	D_{R_T S_T}	&	D_{R_T R_T}  
\end{array} 
\right )
\end{eqnarray*}

\noindent with
\begin{eqnarray*}
&& D_{I I}=2k_{dI}\langle I \rangle, D_{I S_P}=D_{I R_P}=D_{I m}=D_{I S_T}=0, \\
&& D_{I R_T}=D_{S_P I}=0, D_{S_P S_P}=2k_{ap}(\langle S_T\rangle - \langle S_P\rangle)\langle I \rangle, \\
&& D_{S_PR_P}=-k_{k}\langle S_P\rangle(\langle R_T\rangle-\langle R_P\rangle),D_{S_pm}=D_{S_PS_T}=0,\\
&& D_{S_PR_T}=D_{R_PI}=0,D_{R_PS_P}=-k_{k}\langle S_P\rangle(\langle R_T\rangle-\langle R_P\rangle),\\
&& D_{R_PR_P}=2k_{k}\langle S_P\rangle(\langle R_T\rangle-\langle R_P\rangle),D_{R_Pm}=D_{R_PS_T}=0,\\
&& D_{R_PR_T}=D_{mI}=D_{mS_P}=D_{mR_P}=0,D_{mm}=2k_{dm}\langle m\rangle,\\
&& D_{mS_T}=D_{mR_T}=D_{S_TI}=D_{S_TS_P}=D_{S_TR_P}=D_{S_Tm}=0,\\
&& D_{S_TS_T}=2k_{dp}\langle S_T\rangle,D_{S_TR_T}=D_{R_TI}=D_{R_TS_P}=0,\\
&& D_{R_TR_P}=D_{R_Tm}=D_{R_TS_T}=0,D_{R_TR_T}=2k_{dp}\langle R_T\rangle.
\end{eqnarray*}
  



\begin{table*}[ht!]
\caption{
List of kinetics schemes and the values of rate parameters used in the model.
$f(R_P)$= $R_P/(R_P+K)$, with $K=20$ $\mu$M.
} 
\begin{ruledtabular}
\begin{tabular}{lllll||}
{Reaction} & {Kinetics} & {Propensity} & {Rate constant}\\ [0.5ex]
\hline
Synthesis of $I$					&	$\phi \rightarrow I$			&	$k_{sI}$			&	$k_{sI}=0.003\ \mu$Ms$^{-1}$\\
Degradation of $I$					&	$I \rightarrow \phi$			&	$k_{dI}I$			&	$k_{dI}=0.01\ $s$^{-1}$\\
Autophosphorylation of $S$	&	$S+I \rightarrow S_P$		&	$k_{ap}SI$		&	$k_{ap}=1.0\ \mu$M$^{-1}$s$^{-1}$	\\
Autodephosphorylation of $S_P$		&	$S_P \rightarrow S$			&	$k_{adp}S_P$		&	$k_{adp}=0.01$\ s$^{-1}$\\
Phophotransfer from $S_P$ to $R$		&	$S_P+R \rightarrow S+R_P$	&	$k_{k}S_PR$		&	$k_{k}=1.0\ \mu$M$^{-1}$s$^{-1}$\\
Dephosphorylation of $R_P$ by $S$	&	$R_P+S \rightarrow R+S$	&	$k_{p}R_PS$		&	$k_{p}=0.7\ \mu$M$^{-1}$s$^{-1}$\\
$R_P$ mediated transcription			&	$\phi \stackrel{f(R_P)}{\longrightarrow} m$	&	$k_{sm}f(R_P)$	&	$k_{sm}=0.0002\ \mu$Ms$^{-1}$\\
Degradation of mRNA				&	$m \rightarrow \phi$			&	$k_{dm}m$		&	$k_{dm}=0.01\ $s$^{-1}$\\
Translation of $S$ from $m$			&	$m \rightarrow m+S$		&	$k_{ss}m$		&	$k_{ss}=0.02\ $s$^{-1}$\\
Translation of $R$ from $m$			&	$m \rightarrow m+R$		&	$k_{sr}m$			&	$k_{sr}=0.04\ $s$^{-1}$\\
Degradation of $S$					&	$S \rightarrow \phi$			&	$k_{dp}S$		&	$k_{dp}=0.0016\ $s$^{-1}$\\
Degradation of $R$					&	$R \rightarrow \phi$			&	$k_{dp}R$		&	$k_{dp}=0.0016\ $s$^{-1}$\\
Degradation of $S_P$				&	$S_P \rightarrow \phi$		&	$k_{dp}S_P$		&	$k_{dp}=0.0016\ $s$^{-1}$\\
Degradation of $R_P$				&	$R_P \rightarrow \phi$		&	$k_{dp}R_P$		&	$k_{dp}=0.0016\ $s$^{-1}$\\
\end{tabular}
\end{ruledtabular}
\label{table1}
\end{table*}

\end{document}